\documentclass[superscriptaddress,prd,aps,showpacs,nofootinbib,showkeywords,eqsecnum]{revtex4-2}
\usepackage{graphicx,color,amsmath,amsxtra}
\usepackage{epsf}
\usepackage{amssymb}
\usepackage{enumerate}
\usepackage{hhline}
\usepackage{array}
\usepackage{tabularx}
\usepackage{xcolor}
\usepackage[unicode]{hyperref}             
\usepackage{epstopdf}
\begin{document}
\pagestyle{myheadings}
\title{Anisotropic constant-roll inflation for the Dirac-Born-Infeld model}
\author{Duy H. Nguyen}
\email{duy.nguyenhoang@phenikaa-uni.edu.vn}
\affiliation{Phenikaa Institute for Advanced Study, Phenikaa University, Hanoi 12116, Vietnam}
\affiliation{Graduate University of Science and Technology, Vietnam Academy of Science and Technology, Hanoi 11307, Vietnam}
\author{Tuyen M. Pham}
\email{tuyen.phammanh@phenikaa-uni.edu.vn}
\affiliation{Phenikaa Institute for Advanced Study, Phenikaa University, Hanoi 12116, Vietnam}
\affiliation{Graduate University of Science and Technology, Vietnam Academy of Science and Technology, Hanoi 11307, Vietnam}
\author{Tuan Q. Do }
\email{tuan.doquoc@phenikaa-uni.edu.vn}
\affiliation{Phenikaa Institute for Advanced Study, Phenikaa University, Hanoi 12116, Vietnam}
\affiliation{Faculty of Basic Sciences, Phenikaa University, Hanoi 12116, Vietnam}
\date{\today} 
\begin{abstract}
In this paper, we study a non-canonical extension of a supergravity-motivated model acting as a vivid counterexample to the cosmic no-hair conjecture due to its unusual  coupling between scalar and electromagnetic fields. In particular, a canonical scalar field is replaced by the string-inspired Dirac-Born-Infeld one in this extension. As a result, exact anisotropic inflationary solutions for this Dirac-Born-Infeld model are figured out under a constant-roll condition. Furthermore, numerical calculations are performed to verify that these anisotropic constant-roll solutions are indeed attractive during their inflationary phase. 
%Keywords: expanding universes, Bianchi spaces, no-hair theorem, CMB
\end{abstract}
\maketitle
%%%%%%%%%%%%%%%%%%%%%%%%%%%%%%%%%%%%%%%%%
\section{Introduction} \label{intro}
Cosmic inflation has played as one of the leading paradigms in modern cosmology \cite{cosmic-inflation}. This is due to the fact that many theoretical predictions of cosmic inflation have been well confirmed by the cosmic microwave background  radiation (CMB) observations of the Wilkinson Microwave Anisotropy Probe (WMAP) \cite{WMAP} as well as the Planck \cite{Planck} satellites.

 It appears that all standard inflationary models have been investigated under the cosmological principle \cite{Martin:2013tda}, which states that our universe is just simply homogeneous and isotropic on large scales as described by the Friedmann-Lemaitre-Robertson-Walker (FLRW) spacetime \cite{FLRW}. However, testing the validity of this principle seems to be not straightforward \cite{cosmological-principle}.  
Recently, some anomalies of the CMB temperature such as the hemispherical asymmetry and the cold spot, which have been detected by the WMAP and confirmed by the Planck, have challenged the validity of the cosmological principle \cite{Schwarz:2015cma}. Consequently, this issue has received a lot of attention. In particular, a number of mechanisms have been proposed to explain the origin of these anomalies \cite{Schwarz:2015cma}. For example, some papers have appeared with an interesting idea that the mentioned anomalies could  just be instrumental rather than cosmological \cite{Hanson:2010gu}. In particular, the investigations in these papers have pointed out, using either the WMAP or Planck data, that the asymmetric beams could lead to the CMB statistical anisotropy. However, an independent analysis in Ref. \cite{Groeneboom:2009cb} has been performed to show that the asymmetric beams seem to be unimportant. It is worth noting that a recent study has suggested that the Hubble tension might be a smoking gun for a breakdown of the FLRW cosmology \cite{Krishnan:2021dyb}.

All of these motivations lead us to think of a modification of the cosmological principle. One of reasonable possibilities is that the isotropy of spacetime of the early time universe should be slightly broken down \cite{Schwarz:2015cma,Pitrou:2008gk}. Consequently, the spacetime of the early time universe would be described by the Bianchi metrics \cite{bianchi,Pitrou:2008gk}, which are homogeneous but anisotropic, rather than the FLRW one.  If the early time universe was anisotropic, one could ask a related question that would the late time universe still be anisotropic or not. It appears that answering this question is not straightforward task at all. Theoretically, there has existed an important hint based on the so-called cosmic no-hair conjecture proposed by Hawking and his colleagues a few decades ago \cite{cosmic-nohair}. According to this conjecture, a late time state of the universe should obey the cosmological principle, i.e., should be homogeneous and isotropic, regardless of any initial states of the universe, which might have hairs, i.e., inhomogeneity and/or anisotropy. However, it is very complicated to draw a complete proof to this conjecture. Indeed, enormous efforts have been made to prove this conjecture for specific scenarios since the first partial proof by Wald for the Bianchi metrics in the presence of the cosmological constant $\Lambda$ \cite{wald,proof-1,proof-2,proof-3}. It is important to note that if the cosmic no-hair conjecture was valid, it would  only be valid locally, according to the investigations in Refs. \cite{local-1,local-2}.   Along with the theoretical hint, it is worth noting that there has been a recent observational claim that the current universe might be anisotropic \cite{Colin:2018ghy}. If this observation was the case, it would be an important issue directly related to validity of the cosmic no-hair conjecture. 

Along with the proofs, many papers have examined the validity of the cosmic no-hair conjecture within specific inflationary models to see whether it is broken down or not \cite{higher-order-gravity,kaloper,galileon,kao09,MW0,MW}. It turns out that many papers among them have claimed to admit counterexamples to the conjecture. However, most of these claimed counterexamples have been pointed out to be invalid due to their instability \cite{kao09}. Up to now, an anisotropic inflation found in a supergravity-motivated model by Kanno, Soda, and Watanabe (KSW) has been regarded as the first counterexample to the cosmic no-hair conjecture \cite{MW0,MW}. In particular, this anisotropic inflation has been shown to be stable and attractive  due to the existence of supergravity-motivated coupling between scalar and vector fields, $f^2(\phi)F^{\mu\nu}F_{\mu\nu}$ \cite{MW}. Consequently, this model has received a lot of attention. As a result, a number of its extensions have been proposed in order to see whether other counterexamples to the conjecture appear or not \cite{extensions,WFK,Fujita:2018zbr,multi-vector-1,DBI,ghost-condensed,SDBI,Ito:2017bnn}. It motivated a revisiting of the Wald's no-hair theorem for inflationary settings \cite{Maleknejad:2012as}.  Additionally, CMB imprints of the anisotropic inflation within the context of the KSW model as well as its non-canonical extensions have been investigated in Refs. \cite{data,Imprint1,Imprint2,Imprint3,Imprint4}, while primordial gravitational waves in anisotropic inflation have been studied in Ref. \cite{gws}. For other cosmological aspects of the KSW anisotropic inflation, see interesting reviews in Ref. \cite{SD}.

Among non-trivial extensions of the KSW model, there are two interesting approaches, which we are currently interested in. The first approach is  non-canonical extensions \cite{DBI,SDBI,ghost-condensed}, in which a canonical scalar field is replaced by non-canonical ones such as the Dirac-Born-Infeld (DBI) field \cite{DBI}, whose origin can be realized in the D3-brane theory \cite{Silverstein:2003hf}. For interesting cosmological aspects of the DBI inflation, see Refs. \cite{Chen:2005ad,Chen:2006nt,Baumann:2006cd,Spalinski:2007dv,Copeland:2010jt}. As a result, these non-canonical models have been shown to admit stable and attractive anisotropic power-law inflationary solutions, which do violate the validity of the cosmic no-hair conjecture \cite{DBI,SDBI,ghost-condensed}. This result really confirms the leading role of the coupling $f^2(\phi)F^{\mu\nu}F_{\mu\nu}$ in breaking down the validity of the cosmic no-hair conjecture. The last approach is a new type of anisotropic inflation, which has been found within the KSW model under a constant-roll condition \cite{Ito:2017bnn}. 

Basically, the constant-roll condition is based on an assumption that one of slow-roll parameters, $\eta=-\ddot\phi/(H\dot\phi)$, should be constant during an inflationary phase \cite{Motohashi:2014ppa}.  Interestingly, this constant-roll inflation is regarded as an interpolation between the well-known slow-roll inflation and a novel ultra-slow-roll inflation proposed in Ref. \cite{Martin:2012pe}. Additionally, it appears that the well-known power-law inflation associated with an exponential potential of scalar field \cite{Abbott:1984fp} can be re-derived under this constant-roll condition. More interestingly, other types of exact inflationary solutions corresponding to different potentials, which were firstly proposed in Refs. \cite{Barrow:1994nt,Boubekeur:2005zm}, can also be re-derived under this constant-roll condition \cite{Motohashi:2014ppa}. Due to the constancy of $\eta$, ones could expect that the constant-roll inflation might admit theoretical predictions slightly different from that of the slow-roll inflation. In other words, small corrections for the slow-roll inflation might be derived from the constant-roll inflation \cite{Motohashi:2017aob,GalvezGhersi:2018haa}. Consequently, the constant-roll inflation has been investigated extensively in Refs. \cite{Odintsov:2017yud,Anguelova:2017djf,Lin:2019fcz,Motohashi:2019rhu,Karam:2017rpw,Nojiri:2017qvx,Motohashi:2019tyj,Odintsov:2019ahz,Mohammadi:2018wfk,Mohammadi:2018zkf,Antoniadis:2020dfq,Gao:2020cvb,Guerrero:2020lng,Sadeghi:2021egp,Shokri:2021rhy}. It appears that the constant-roll condition has been applied to a number of cosmological models such as the non-minimal Coleman-Weinberg \cite{Karam:2017rpw}, $F(R)$ gravity \cite{Nojiri:2017qvx}, scalar-tensor gravity  \cite{Motohashi:2019tyj}, $k$-inflation \cite{Odintsov:2019ahz}, non-canonical scalar field \cite{Mohammadi:2018wfk}, DBI \cite{Mohammadi:2018zkf}, Palatini-$R^2$ gravity \cite{Antoniadis:2020dfq}, Gauss-Bonnet gravity \cite{Gao:2020cvb}, multi scalar fields \cite{Guerrero:2020lng}, non-commutative gravity \cite{Sadeghi:2021egp}, and $f(\phi)R$ gravity \cite{Shokri:2021rhy} models to figure out either novel (exact) inflationary solutions or cosmological consequences. 

Hence, it would be interesting if we were able to figure out anisotropic inflation under the constant-roll condition within the KSW model. It is worth noting that the study in Ref. \cite{Ito:2017bnn} has pointed out that it is possible to have such anisotropic inflation for the KSW model of canonical scalar field. Furthermore, it has been shown that the anisotropic constant-roll inflation is really attractive \cite{Ito:2017bnn}. All of these motivations lead us to propose to investigate in this paper whether the DBI extension of the KSW model \cite{DBI} admits anisotropic constant-roll inflation. As a result, we will point out the existence of anisotropic DBI constant-roll inflationary solutions. Furthermore, we will numerically confirm the attractive property of these solutions. 

As a result, the present paper will be organized as follows: (i) A brief introduction has been written in Sec. \ref{intro}. (ii) Basic setup of the DBI model will be presented in Sec. \ref{chap2}. (iii) Anisotropic constant-roll inflationary solutions will be solved in Sec. \ref{chap3}. (iv) Attractor property of the obtained solutions will be shown in Sec. \ref{chap4}. (v) Finally, concluding remarks will be given in Sec. \ref{final}.
%%%%%%%%%%%%%%%%%%
\section{Dirac-Born-Infeld model} \label{chap2}
In this paper, we would like to study a non-canonical extension of the KSW model \cite{DBI}, in which a canonical scalar field is replaced by the non-canonical Dirac-Born-Infeld (DBI) field \cite{Silverstein:2003hf} as follows
\begin{align} \label{action}
S=\int d^4x\sqrt{-g}\left[\frac{R}{2}+\frac{1}
{{f\left( \phi  \right)}}\frac{\gamma-1}{\gamma} -V\left( \phi  \right)-\frac{1}{4}h^2(\phi)F_{\mu\nu}F^{\mu\nu}\right],
\end{align}
where the reduced Planck mass $M_p$ is set to be one for convenience. In addition, $V(\phi)$ is the potential of scalar field $\phi$ and $F_{\mu\nu}\equiv\partial_\mu A_\nu-\partial_\nu A_\mu$ is the field strength of the vector field $A_\mu$.  It is noted that the notation $\gamma$ defined as
\cite{Silverstein:2003hf}
\begin{equation}
\gamma  \equiv \frac{1}{ \sqrt {1 + f (\phi) \partial _\mu  \phi \partial ^\mu  \phi}},
\end{equation}
plays as the Lorentz factor characterizing the motion of the D3-brane \cite{Silverstein:2003hf}. It is clear that $\gamma \geq 1$ for non-negative $f(\phi)$. 
 The last ingredient in the action \eqref{action} is a non-trivial coupling function $h(\phi)$, whose existence breaks down the conformal invariance of the Maxwell field. It should be noted that the existence of $f(\phi)$ in the definition of $\gamma$ leads us to modify the original KSW coupling $f^2(\phi)F_{\mu\nu}F^{\mu\nu}$ to $h^2(\phi)F_{\mu\nu}F^{\mu\nu}$ in order to avoid any misunderstanding \cite{DBI}. It is clear that if we take the canonical limit, i.e., $f \to 0$, then the DBI model will reduce to the original KSW model.

As a result, a stable and attractive Bianchi type I power-law inflation has been found in this DBI model \cite{DBI}. This implies that the non-canonical property of the DBI field does not affect on the breaking of the cosmic no-hair conjecture within the KSW model. In other words, the coupling between the scalar and vector field, $h^2(\phi)F_{\mu\nu}F^{\mu\nu}$, does play the leading role in violating the cosmic no-hair conjecture. It is worth noting that the CMB imprints of non-canonical anisotropic model have been investigated recently in Ref. \cite{Imprint4}. 

As a result, the corresponding field equations of the DBI can be derived as
 \begin{align} \label{Eintein-field-equation}
  \left( {R_{\mu \nu }  - \frac{1}
{2}Rg_{\mu \nu } } \right) - \gamma \partial _\mu  \phi \partial _\nu  \phi  
+ g_{\mu \nu } \left[ {\frac{1}
{{f}}\left( {\frac{1}
{\gamma }-1} \right) + V+ \frac{1}
{4} h^2 F^{\rho \sigma } F_{\rho \sigma } } \right]   - h^2 F_{\mu \gamma } F_\nu{} ^\gamma  & = 0, \\
   \label{scalar-field-equation}
   \partial_\mu(\gamma\sqrt{-g}\,\partial^\mu\phi)-\sqrt{-g}\left[\frac{ f'}{2f}\frac{(\gamma-1)\gamma}{\gamma+1}\partial_\mu\phi\partial^\mu\phi+V'+\frac{hh'}{2}F_{\mu\nu}F^{\mu\nu}\right]&=0 ,\\
  \label{vector-field-equation}
 \partial _\mu  \left[ {\sqrt { - g} \left( {h^2 F^{\mu \nu } } \right)} \right] & = 0,
\end{align}
where $V'\equiv dV/d\phi$ and so on.
In this paper, similar to the previous work in Ref. \cite{Ito:2017bnn}, we will consider the following Bianchi type I metric, which is homogeneous but anisotropic, 
\begin{align}
ds^2=-dt^2+a^2(t)\left[b^{-4}(t)dx^2+b^2(t)\left(dy^2+dz^2\right)\right],
\end{align}
here $a(t)$ acts as an isotropic scale factor, while $b(t)$ plays as a deviation from isotropy. In addition, we will assume that the scalar field is homogeneous, i.e., $\phi=\phi(t)$, while the vector field $A_\mu$ takes the following configuration as $A_\mu=\left(0,A_x(t),0,0\right)$ in order to be compatible with the Bianchi type I metric. As a result, a non-trivial solution to the field equation of vector field \eqref{vector-field-equation} can be solved to be
\begin{align}
\dot{A}_x=p_A h^{-2}a^{-1}b^{-4},
\end{align}
where $p_A$ is an integration constant. Thanks to this solution, the corresponding field equation for the scalar field \eqref{scalar-field-equation} now reduces to
\begin{align}\label{DBI scalar field equation}
\ddot{\phi}=-\frac{3H_a}{\gamma^2}\dot{\phi}-\frac{V'}{\gamma^3}-\frac{f'}{2f}\frac{(\gamma+2)(\gamma-1)}{(\gamma+1)\gamma}\dot{\phi}^2+\frac{h^{-3}h'}{\gamma^3}p_A^2a^{-4}b^{-4}, 
\end{align}
where the corresponding Lorentz factor now becomes
\begin{align}\label{definition of gamma}
\gamma = \frac{1}{\sqrt{1-f\dot{\phi}^2}}.
\end{align}
Here $\dot{A}_x \equiv dA_x/dt$, while $H_a\equiv{\dot{a}}/{a}$ and $H_b\equiv{\dot{b}}/{b}$ are two Hubble parameters. As a result,  the corresponding non-vanishing components of the Einstein equation \eqref{Eintein-field-equation} turn out to be 
\begin{align}
\label{1st DBI Einstein equation}
H_a^2-H_b^2&=\frac{1}{3}\left[\frac{\gamma^2}{\gamma+1}\dot{\phi}^2+V+\frac{p_A^2}{2}h^{-2}a^{-4}b^{-4}\right], \\
\label{2nd DBI Einstein equation}
\dot{H}_a&=-3H_b^2-\frac{\gamma}{2}\dot{\phi}^2-\frac{p_A^2}{3}h^{-2}a^{-4}b^{-4},\\ 
\label{3rd DBI Einstein equation}
\dot{H}_b&=-3H_aH_b+\frac{p_A^2}{3}h^{-2}a^{-4}b^{-4}.
 \end{align}
It appears that Eq. \eqref{1st DBI Einstein equation} is the Friedmann equation acting as a constraint equation, while Eqs. \eqref{2nd DBI Einstein equation} and \eqref{3rd DBI Einstein equation} play as the evolution equation of the scale factors $a(t)$ and $b(t)$, respectively.  Up to now, we have derived all the field equations of the DBI model, that are Eqs. \eqref{DBI scalar field equation}, \eqref{1st DBI Einstein equation}, \eqref{2nd DBI Einstein equation}, and \eqref{3rd DBI Einstein equation}. Once again, it appears that if we take the canonical limit $\gamma \to 1$, or equivalently $f\to 0$, then all these field equations will reduce to that derived in the KSW model for canonical scalar field \cite{Ito:2017bnn}. Additionally, if $a(t)$ and $b(t)$ both take exponential forms as $a(t)=\exp[\alpha(t)]$ and $b(t)=\exp[\sigma(t)]$ then all these field equations will become that derived in Ref. \cite{DBI} for the DBI field.
%%%%%%%%%%%%%%%%%%%%% 
\section{Solutions under constant-roll conditions} \label{chap3}
\subsection{Constant-roll conditions}
We recall that in the case of canonical inflation, $\gamma = 1$, Eq. \eqref{DBI scalar field equation} reduces to
\begin{align} \label{canonical}
\ddot{\phi}=-3H_a\dot{\phi}-V'+h^{-3}h'p_A^2a^{-4}b^{-4}.
\end{align}
It is noted that the slow-roll inflation is characterized by some slow-roll parameters, which are normally constrained by the observational data of WMAP or Planck \cite{WMAP,Planck}. One of them is defined as \cite{Motohashi:2014ppa}
\begin{align} \label{def-of-eta}
\eta(t) \equiv-\frac{\ddot{\phi}}{H_a\dot{\phi}}.
\end{align}
Usually, the value of $\eta$ should be much smaller than one within the context of slow-roll inflation. In the constant-roll inflation \cite{Motohashi:2014ppa}, $\eta$ has been assumed to be constant during the inflationary phase, i.e.,
\begin{equation} \label{canonical constant-roll condition}
-\frac{\ddot{\phi}}{H_a\dot{\phi}} = \hat\beta,
\end{equation} 
with $\hat\beta$ is a constant. Note that $\hat\beta$ is related to other definitions of constant-roll constant  in Refs. \cite{Ito:2017bnn,Motohashi:2014ppa,Motohashi:2017aob} such as $\hat\beta=-\beta =3+\alpha$ with $\alpha $ and $\beta$ are also constants. So $\alpha \to -3$ will lead to $\hat\beta \to 0$. As a result, along with the power-law inflation \cite{Abbott:1984fp} other types of inflation corresponding to different potentials can be obtained under this constant-roll assumption \cite{Motohashi:2014ppa}. More interestingly, the similar result also happens in the KSW model \cite{Ito:2017bnn}. 

It has been shown in Ref. \cite{Mohammadi:2018zkf} that the slow-roll parameter $\eta$ could be modified for the DBI scalar field. To see how to modify $\eta$, let us rewrite Eq. \eqref{DBI scalar field equation} as 
\begin{align}
\frac{d}{dt} \left(\gamma\dot{\phi} \right)=- 3H_a \left(\gamma\dot{\phi} \right)-\frac{f'}{2f^2}\left(2-\frac{1}{\gamma}-\gamma\right)-V'+ p_A^2 h^{-3}h'a^{-4}b^{-4}. 
\end{align}
By comparing this equation with that of canonical scalar field \eqref{canonical}, we can arrive at the modified definition of $\eta$ for the DBI model such as 
 \begin{align} 
 \eta_{\rm DBI}(t)= -\frac{1}{H_a(\gamma\dot{\phi})} \frac{d (\gamma\dot{\phi})}{dt},
\end{align}
which is nothing but that introduced in a recent paper \cite{Mohammadi:2018zkf}. However, it turns out that $\eta_{\rm DBI} = \eta$ if $\gamma = {\text {constant}}$.  It should be noted that the condition $\gamma =\text{constant}$ is necessary for obtaining the corresponding power-law inflation to the DBI model as shown in Refs. \cite{DBI,Spalinski:2007dv}. Hence, one can expect the same thing for the DBI constant-roll inflation, which involves the power-law inflation as its subclass. Hence, we will assume from now on the constant Lorentz factor condition,
\begin{align}
\label{constant speed of sound condition}
\gamma=\gamma_0, 
\end{align}
with $\gamma_0 \geq 1$ is a constant. Consequently, the constant-roll condition for the DBI model is now given by
\begin{equation} \label{DBI constant-roll condition}
\eta_{\rm DBI} = \hat\beta.
\end{equation}
Additionally, we will impose the constant anisotropy condition from Ref. \cite{Ito:2017bnn},
\begin{align}
\label{constant anisotropy condition}
\frac{H_b}{H_a}=n,
\end{align}
with $n$ is also a constant, for seeking anisotropic constant-roll solutions. It appears that $|n|\ll 1$ is a sufficient constraint for viable anisotropic constant-roll inflation.

In the following steps, we will employ the method presented in Ref. \cite{Ito:2017bnn} with some modifications necessary for the DBI model. 
More specifically, the strategy of seeking solutions is as follows \cite{Ito:2017bnn}: (i) First, we use three conditions shown in Eqs. \eqref{constant speed of sound condition}, \eqref{DBI constant-roll condition}, and \eqref{constant anisotropy condition} to derive the corresponding forms of $V(\phi)$, $h(\phi)$, and $f(\phi)$. (ii) Then, we put back  the derived forms of $V(\phi)$, $h(\phi)$, and $f(\phi)$ to solve numerically the field equations \eqref{DBI scalar field equation}, \eqref{1st DBI Einstein equation}, \eqref{2nd DBI Einstein equation}, and \eqref{3rd DBI Einstein equation} for different initial conditions along with some values of $\gamma_0$. The attractor properties such as $\eta_{\rm DBI} (t) \to  \hat\beta$, $H_b(t)/H_a(t) \to n$, and $\gamma(t) \to \gamma_0$ after some e-folds during the inflationary phase, will therefore be confirmed by numerical calculations accordingly. 
%%%%%%%%%%%%%%%%
\subsection{Solutions}
As a result, we get, from Eqs. \eqref{2nd DBI Einstein equation}, \eqref{3rd DBI Einstein equation}, and \eqref{constant anisotropy condition}, an equation of $H_a$ such as
\begin{align}
(1+n)\dot{H}_a=-3n(1+n)H_a^2-\frac{\gamma_0}{2}\dot{\phi}^2,
\end{align}
which can be rewritten, by regarding $H_a$ as a function of $\phi$, as follows
\begin{align}
\label{quadratic equation of dphi/dt}
\frac{\gamma_0}{2}\dot{\phi}^2+(1+n)H_a'\dot{\phi}+3n(1+n)H_a^2=0. 
\end{align}
It turns out that Eq. \eqref{quadratic equation of dphi/dt} is a quadratic equation of $\dot{\phi}$ and can therefore  be easily solved to give the corresponding non-trivial solutions, 
\begin{align}
\label{soloution for dphi/dt}
\gamma_0\dot{\phi}=-(1+n)H_a'\pm\sqrt{\left[(1+n)H_a'\right]^2-6\gamma _0n(1+n)H_a^2}.
\end{align}
Furthermore, by differentiating these solutions with respect to $t$, we get
\begin{align}
\label{equation for H_a in DBI case}
-\hat\beta\gamma_0 H_a=-(1+n)H_a''\pm\frac{\sqrt{1+n}\left[(1+n)H_a'H_a''-6n\gamma_0 H_aH_a'\right]}{\sqrt{(1+n)(H_a')^2-6n\gamma_0 H_a^2}},
\end{align}
with the help of Eq. \eqref{DBI constant-roll condition}. Here $H''_a \equiv d^2H_a/d\phi^2$. It is noted that an interesting solution for $H_a$ has been solved in Ref. \cite{Ito:2017bnn} for the canonical model with $\gamma_0=1$  to be
\begin{align}
H_a=C_1\exp\left(\sqrt{\frac{6\hat\beta}{6+\hat\beta}}\phi\right)+C_2\exp\left(-\sqrt{\frac{6\hat\beta}{6+\hat\beta}} \phi\right)
\end{align}
along with the following important relation between two parameters $n$ and $\hat\beta$ \cite{Ito:2017bnn},
\begin{align}
\label{relation of n and beta}
n=\frac{\hat\beta}{6}.
\end{align}
Here, $C_1$ and $C_2$ are integration constants. It should be noted that this solution is different from that derived for an isotropic inflation with $n=0$ in Ref. \cite{Motohashi:2014ppa}.  Similarly, we are able to figure out, thanks to the relation \eqref{relation of n and beta}, a generalized solution for $H_a$ in the DBI model  such as
\begin{align}\label{general}
H_a=C_1\exp\left(\sqrt{\frac{6\gamma_0\hat\beta}{6+\hat\beta}}\phi\right)+C_2\exp\left(-\sqrt{\frac{6\gamma_0\hat\beta}{6+\hat\beta}}\phi\right).
\end{align}
It turns out that different values of $C_1$ and $C_2$ will lead to different types of inflation. It is interesting to note that if we set either $C_1=0$ or $C_2=0$, the corresponding anisotropic power-law inflation \cite{DBI} can be derived accordingly. Note again that the condition $\gamma=\gamma_0$ has been shown to be attractive  in the DBI model for anisotropic power-law inflation \cite{DBI}. Furthermore, following Refs. \cite{Ito:2017bnn,Motohashi:2014ppa} we will focus on three special solutions derived from the general one \eqref{general},
\begin{align}
\label{Ha exp}
H_a&=M\exp\left(\sqrt{\frac{6\gamma_0\hat\beta}{6+\hat\beta}}\phi\right),\\
\label{Ha cosh}
H_a&=M\cosh\left(\sqrt{\frac{6\gamma_0\hat\beta}{6+\hat\beta}}\phi\right),\\
\label{Ha sinh}
H_a&=M\sinh\left(\sqrt{\frac{6\gamma_0\hat\beta}{6+\hat\beta}}\phi\right),
\end{align}
where $M$ is an integration constant determining the amplitude of the power spectrum of the curvature perturbation \cite{Motohashi:2014ppa}. As a result, the first solution \eqref{Ha exp} is obtained by setting $C_1=M$ and $C_2 =0$, while the second one \eqref{Ha cosh} is derived by setting $C_1 =C_2 =M/2$. Additionally, the last solution \eqref{Ha sinh} corresponds to $C_1=M/2$ and $C_2=-M/2$.   It is noted that an unlisted solution corresponding to $C_1=0$ and $C_2=M$, i.e., $H_a=M\exp\left(-\sqrt{\frac{6\gamma_0\hat\beta}{6+\hat\beta}}\phi\right)$, can be obtained from the solution in Eq. \eqref{Ha exp} by redefining $\phi\rightarrow -\phi$ \cite{Motohashi:2014ppa}. It is clear that $\hat\beta$ should be positive definite for the first solution shown in Eq. \eqref{Ha exp}, while $\hat\beta$ can be negative definite for the last two solutions shown in Eqs. \eqref{Ha cosh} and \eqref{Ha sinh} \cite{Motohashi:2014ppa}. However, the relation \eqref{relation of n and beta} implies that $|\hat\beta|$ should be much smaller than one such that $|H_b|\ll H_a$ in order to be consistent with the observations. It appears that these solutions will recover that derived in Ref. \cite{Ito:2017bnn} for the canonical scalar field once taking  a limit $\gamma_0\to 1$. However, these solutions as well as the solutions found in Ref. \cite{Ito:2017bnn} cannot reduce to that derived  in Ref. \cite{Motohashi:2014ppa} for an isotropic inflation under a limit $n\to0$. This is because these solutions have been derived using a special relation \eqref{relation of n and beta}, which was firstly adopted in Ref. \cite{Ito:2017bnn} for the anisotropic inflation.
%%%%%%%%%%%
 \subsubsection{First solution $H_a=M\exp\left(\sqrt{\frac{6\gamma_0\hat\beta}{6+\hat\beta}}\phi\right)$}\label{case-1}
We first consider the solution shown in Eq. \eqref{Ha exp}. Thanks to this solution Eq. \eqref{soloution for dphi/dt} will be reduced to an useful relation
\begin{align}
\label{dphi/di in terms of phi in the case Ha exp}
\dot{\phi}= -M \sqrt{\frac{\hat\beta(6+\hat\beta)}{6\gamma_0}} \exp\left(\sqrt{\frac{6\gamma_0\hat\beta}{6+\hat\beta}}\phi\right),
\end{align}
which helps us to figure out the corresponding values (up to a constant) of the scale factors  in terms of $\phi(t)$ such as
\begin{align}
a(t)&\propto\exp\left(-\sqrt{\frac{6\gamma_0}{\hat\beta(6+\hat\beta)}}\phi (t)\right),\\
b(t)&\propto\exp\left(-\frac{\hat\beta}{6}\sqrt{\frac{6\gamma_0}{\hat\beta(6+\hat\beta)}}\phi (t)\right).
\end{align}
Consequently, the corresponding potential $V(\phi)$ and coupling function $h(\phi)$ can be defined, according to Eqs. \eqref{1st DBI Einstein equation} and \eqref{3rd DBI Einstein equation}, to be
\begin{align}
\label{explicit form of V in Ha exp}
V(\phi)=\frac{3\gamma_0 \left(12-7\hat\beta \right)+2\hat\beta^2-9\hat\beta+36}{12(\gamma_0+1)}M^2\exp\left(2\sqrt{\frac{6\gamma_0 \hat\beta}{6+\hat\beta}}\phi\right)
\end{align}
and, up to a constant,
\begin{align}
\label{explicit form of h in Ha exp}
h (\phi) \propto\exp\left(\sqrt{\frac{6\gamma_0\hat\beta}{6+\hat\beta}}\frac{6-2\hat\beta}{3\hat\beta}\phi\right),
\end{align}
respectively. From the definition of $\gamma$ shown in Eq. \eqref{definition of gamma}, we obtain the corresponding definition of $f(\phi)$ as
\begin{align}
\label{explicit form of f in Ha exp}
f(\phi)=\frac{6(\gamma_0^2-1)}{M^2\gamma_0\hat\beta(6+\hat\beta)}\exp\left(-2\sqrt{\frac{6\gamma_0 \hat\beta}{6+\hat\beta}}\;\phi\right),
\end{align}
with the help of the solution \eqref{dphi/di in terms of phi in the case Ha exp}. In conclusion, for the solution \eqref{Ha exp} under the constant-roll condition, the corresponding functions $f(\phi)$ and $h(\phi)$ as well as the corresponding potential $V(\phi)$ all appear to be exponential functions of $\phi$. As a result, Eq. \eqref{dphi/di in terms of phi in the case Ha exp} can be solved to give the following solution,
\begin{align}
\phi(t)=-\sqrt{\frac{6+\hat\beta}{6\hat\beta\gamma_0}}\log \left(M\hat\beta t \right),
\end{align}
which leads the scale factors $a(t)$ and $b(t)$ to the following power-law expansion, up to a constant, 
\begin{align} \label{scale-factor-1}
a \propto t^\frac{1}{\hat\beta},~b\propto t^\frac{1}{6},
\end{align}
respectively. Hence, it turns out that $0<\hat\beta \ll 1$ is a sufficient condition in order to have anisotropic inflation.  This power-law solution is nothing but that derived in Ref. \cite{DBI} for the DBI model.
%%%%%%%%%%%%%%%%%%%%%%%%%%%%%%%%
\subsubsection{Second solution $H_a=M\cosh\left(\sqrt{\frac{6\gamma_0\hat\beta}{6+\hat\beta}}\phi\right)$} \label{case-2}
As a result, substituting the second solution \eqref{Ha cosh} into Eq. \eqref{soloution for dphi/dt} indicates that $\hat{\beta}$ must be negative definite in order to make $\dot{\phi}$ real definite. However, it is confirmed in Ref. \cite{Ito:2017bnn} that this solution is not an attractor for the canonical scalar field with $\gamma_0=1$. Additionally, this is also the case for the DBI model. Therefore, we will ignore this solution from now on. It should be noted that  the ``cosh" type solution of Hubble constant $H$ has been shown to be physical for the isotropic inflation \cite{Motohashi:2014ppa}. This is indeed a different point between the isotropic constant-roll inflation found in Ref. \cite{Motohashi:2014ppa} and the anisotropic constant-roll inflation found in Ref. \cite{Ito:2017bnn} as well as in this paper.
%%%%%%%%%%%%%%%%%%%%%%%%%%%%%%%%
\subsubsection{Third solution $H_a=M\sinh\left(\sqrt{\frac{6\gamma_0\hat\beta}{6+\hat\beta}}\phi\right)$} \label{case-3}
Now, we consider the last solution shown in Eq. \eqref{Ha sinh}. As a result, this solution leads Eq. \eqref{soloution for dphi/dt} to the corresponding useful relation,
\begin{align}
\label{dphi/di in terms of phi in the case Ha sinh}
\dot{\phi}=M\sqrt{\frac{\hat\beta(6+\hat\beta)}{6\gamma_0}}\left[\pm1-\cosh\left(\sqrt{\frac{6\gamma_0\hat\beta}{6+\hat\beta}}\phi\right)\right],
\end{align}
which helps us to define the corresponding values, up to a constant, of the scale factors to be
\begin{align}
a(t)&\propto\left[\mp1+\cosh\left(\sqrt{\frac{6\gamma_0\hat\beta}{6+\hat\beta}}\phi(t)\right)\right]^{-\frac{1}{\hat\beta}},\\
b(t)&\propto\left[\mp1+\cosh\left(\sqrt{\frac{6\gamma_0\hat\beta}{6+\hat\beta}}\phi(t)\right)\right]^{-\frac{1}{6}}.
\end{align}
Consequently, the corresponding definitions of $V(\phi)$, $h(\phi)$, and $f(\phi)$ turn out to be
\begin{align}
\label{explicit form of V in Ha sinh}
V(\phi)=&-\frac{M^2}{24(\gamma_0+1)}\left\{ \gamma_0 \left( 2\hat\beta^2+27\hat\beta+36\right)-4\hat\beta^2-9\hat\beta+36\right. \nonumber\\
& \left.+\left[3 \gamma_0 \left(7\hat\beta-12\right)-2\hat\beta^2+9\hat\beta-36\right]\cosh\left(2\sqrt{\frac{6\gamma_0\hat\beta}{6+\hat\beta}}\phi\right)\right.\nonumber\\
&\left.\mp 2 \left[ \gamma_0 \left(\hat\beta+24\right)-3\hat\beta \right]\hat\beta\cosh\left(\sqrt{\frac{6\gamma_0\hat\beta}{6+\hat\beta}}\phi\right)\right\},\\
\label{explicit form of h in Ha sinh}
h(\phi)\propto &~ \frac{\left[\mp1+\cosh\left(\sqrt{\frac{6\gamma_0\hat\beta}{6+\hat\beta}}\;\phi\right)\right]^{\frac{6+\hat\beta}{3\hat\beta}}}{\sqrt{-(\hat\beta+3)-(\hat\beta-3)\cosh\left(2\sqrt{\frac{6\gamma_0\hat\beta}{6+\hat\beta}}\phi\right)\pm2\hat\beta\cosh\left(\sqrt{\frac{6\gamma_0\hat\beta}{6+\hat\beta}}\phi\right)}},\\
\label{explicit form of f in Ha sinh}
f(\phi)=&~ \frac{6(\gamma_0^2-1)}{M^2\gamma_0\hat\beta(6+\hat\beta)}\left[\pm1-\cosh\left(\sqrt{\frac{6\gamma_0\hat\beta}{6+\hat\beta}}\phi\right)\right]^{-2},
\end{align}
respectively.  Here, we have only considered $\hat{\beta}>0$. For $\hat{\beta}<0$, it appears that the solution \eqref{Ha sinh} is similar, up to a field redefinition, to the solution \eqref{Ha cosh}  \cite{Motohashi:2014ppa}, and therefore is neglected.  It turns out that these solutions will be reduced to that derived for the canonical scalar field in Ref. \cite{Ito:2017bnn} by taking the limit $\gamma_0 \to 1$. As a result, once  Eq. \eqref{dphi/di in terms of phi in the case Ha sinh} is solved to give the solution of $\phi$, we can easily determine the related scale factors $a(t)$ and $b(t)$ as well as the functions $f(\phi)$, $h(\phi)$, and $V(\phi)$. 

\begin{figure}[hbtp]
 {\includegraphics[scale=0.4]{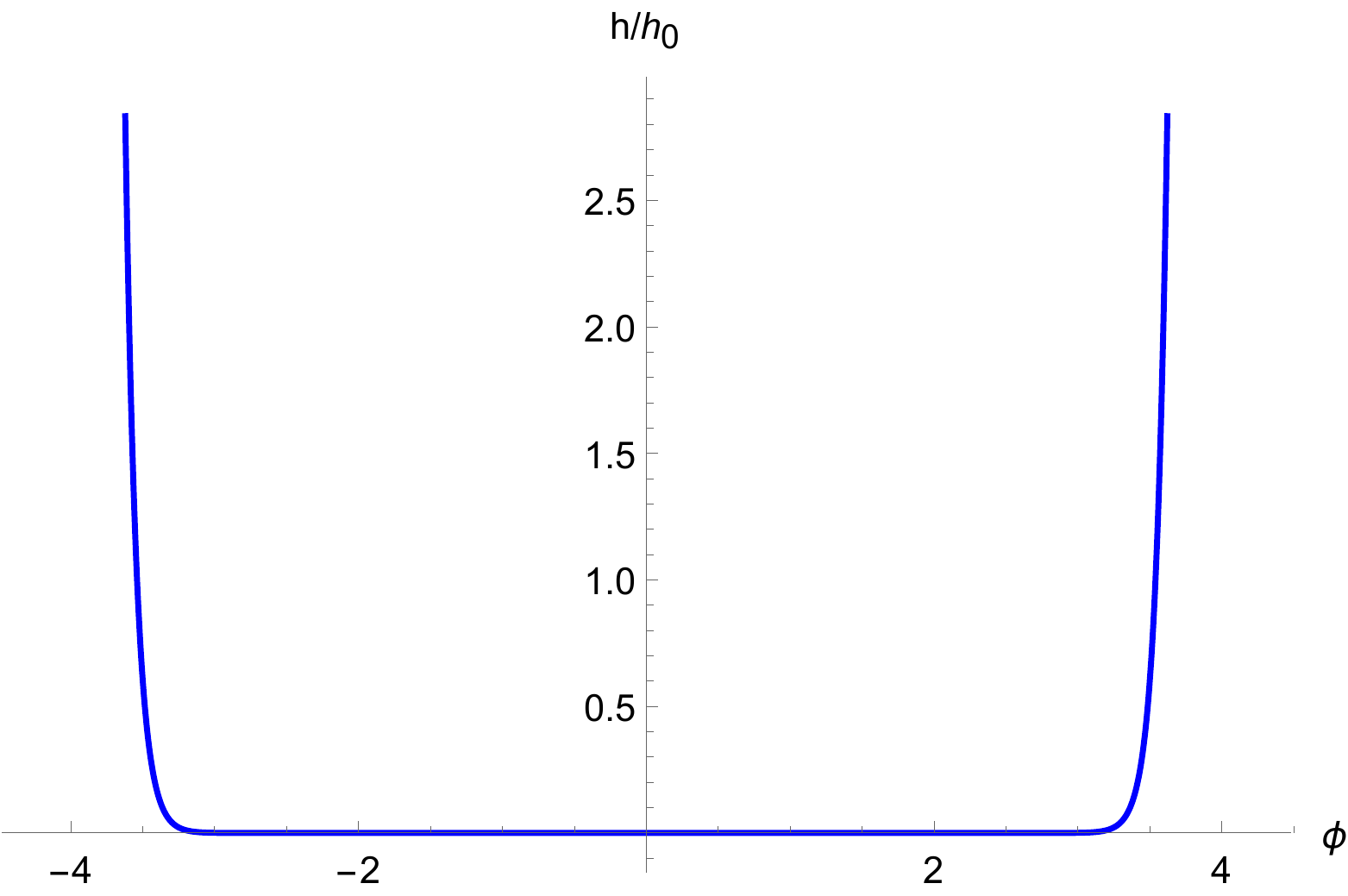}}\quad\
  {\includegraphics[scale=0.4]{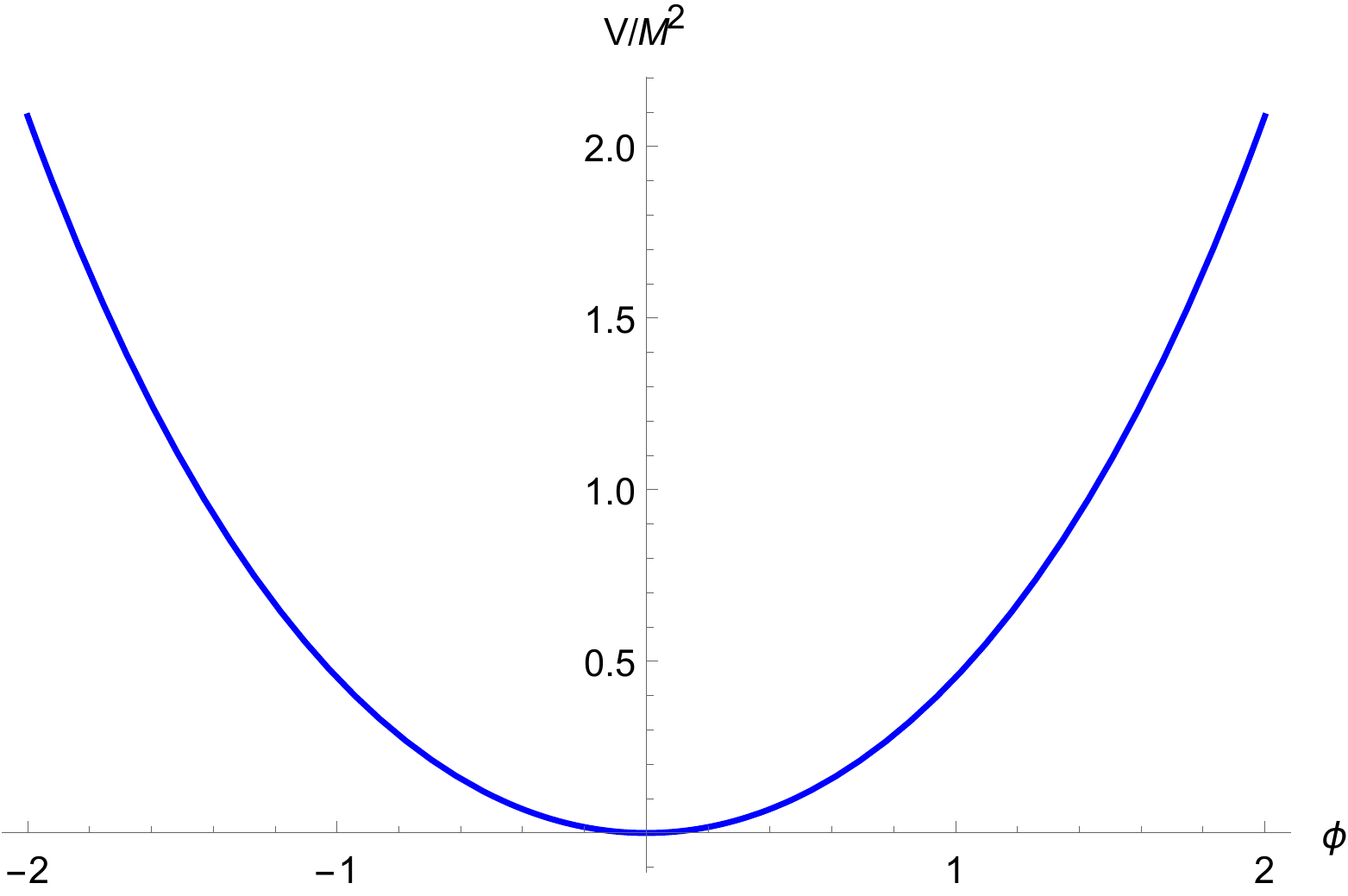}}\\
   \centering
 \caption{Behavior of $h(\phi)$ (left) and $V(\phi)$ (right) associated with $\phi_+(t)$ for $\hat{\beta}=0.1$ and $\gamma_0=1.5$. Here, $h_0$ is an integration constant. This plot shows that $h(\phi)$ is smaller than one when $\phi$ is close to zero. In addition, it also shows that $V(\phi)$ is not negative definite for all values of $\phi$.}
  \label{V for upper sign}
\end{figure}

As a result, a non-trivial solution of Eq.  \eqref{dphi/di in terms of phi in the case Ha sinh} for the upper $``+"$ sign is given by
\begin{equation} \label{phi-upper-sign}
\phi(t)=\phi_+(t)=\sqrt{\frac{2(6+\hat\beta)}{3\hat\beta\gamma_0}}\text{arccoth}\left( M\hat\beta t\right),
\end{equation}
which leads to the corresponding scale factors given by
\begin{align} \label{scale-a-upper}
a(t)&\propto\left(M^2\hat\beta^2 t^2-1 \right)^{\frac{1}{\hat\beta}},\\
\label{scale-b-upper}
b(t)&\propto\left(M^2\hat\beta^2 t^2-1 \right)^{\frac{1}{6}}.
\end{align}
It appears that $a(t)$, $b(t)$, and $\phi(t)$ are all real definite for $M^2\hat\beta^2 t^2>1$. However, it turns out that only the region $M\hat\beta t>1$ is suitable for inflationary solutions since the other region $M\hat\beta t<-1$ leads to $\dot{a}(t)<0$. Moreover, it is clear that inflation will happen if $0<\hat\beta \ll 1$. Hence, $M$ should be positive definite in this case. Additionally, it is noted that this solution in the canonical limit, $\gamma_0=1$, has been claimed to suffer a strong coupling problem around $\phi=0$, i.e., $h(\phi \simeq 0)<1$ \cite{Ito:2017bnn}. Consequently, this problem leads to a requirement that the inflation  needs to finish before $\phi$ becomes zero. It is straightforward to check that this result is also the case for $\gamma_0 >1$ (see Fig.  \ref{V for upper sign} for a demonstration).

\begin{figure}[hbtp]
 {\includegraphics[scale=0.40]{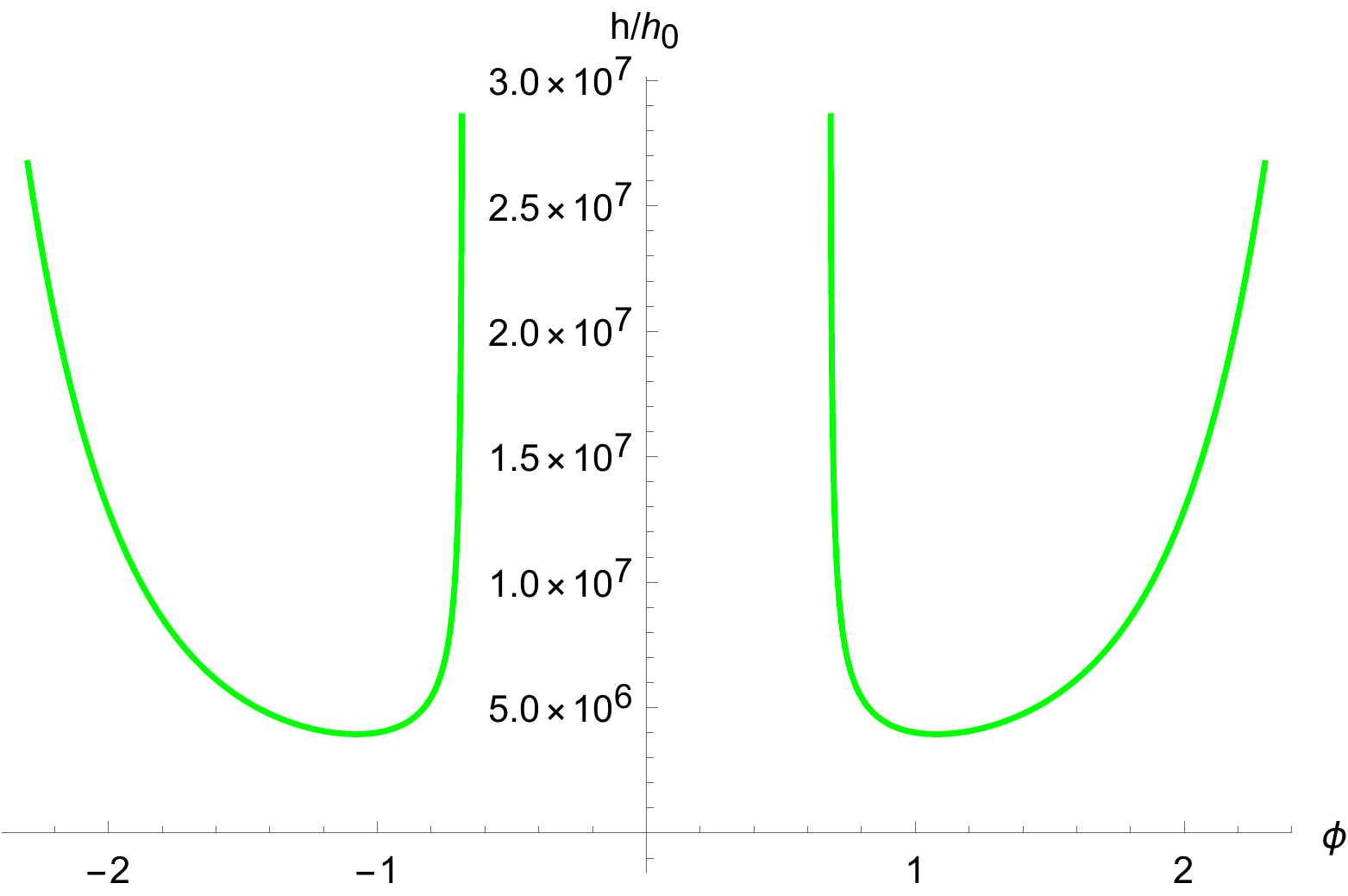}}\quad\
 {\includegraphics[scale=0.40]{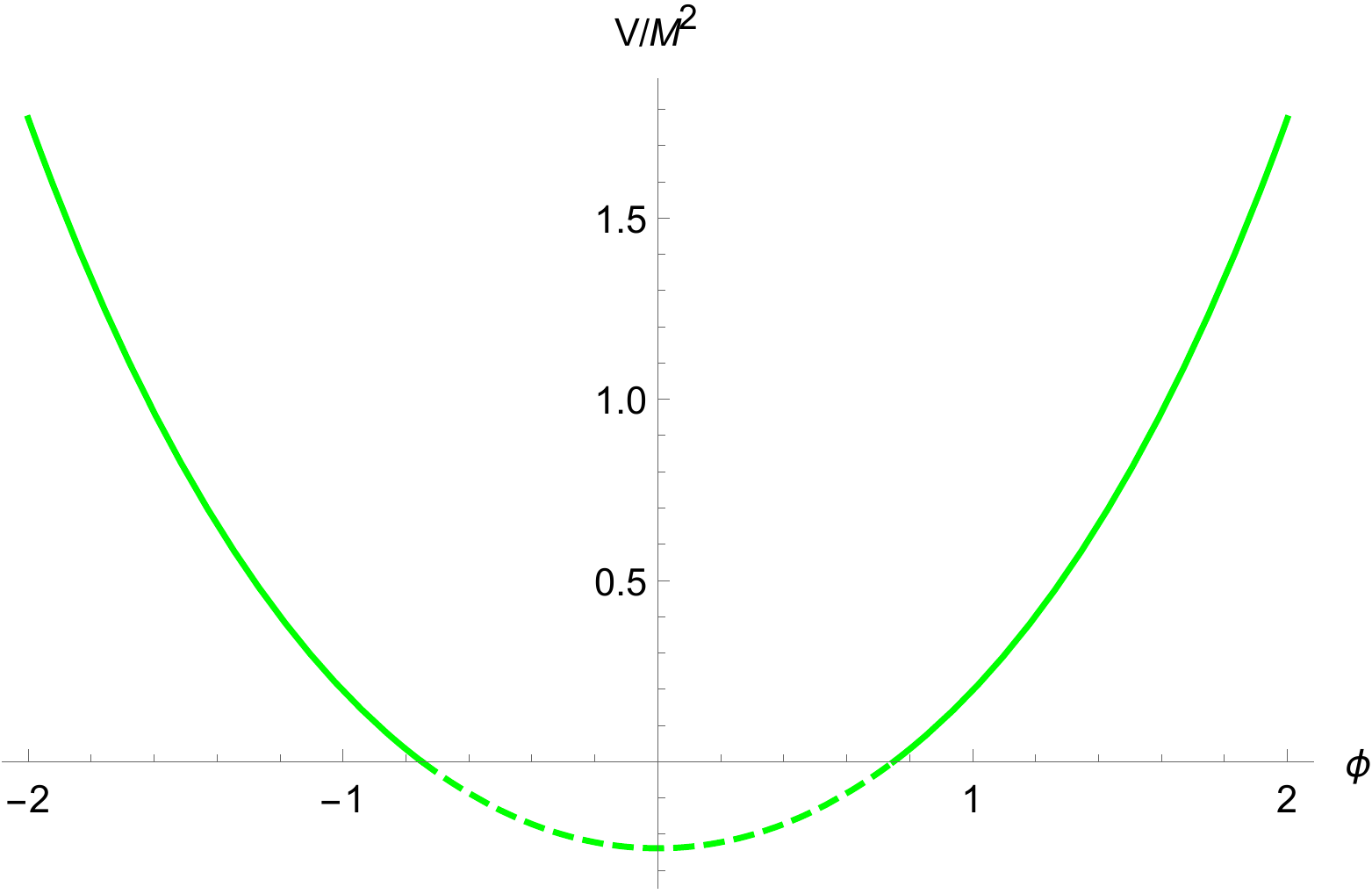}}
   \centering
 \caption{Behavior of $h(\phi)$ (left) and $V(\phi)$ (right) associated with $\phi_-(t)$ for $\hat{\beta}=0.1$ and $\gamma_0=1.5$. Here, $h_0$ is an integration constant. This plot shows  that $h(\phi)$ is always larger than one. In addition, it also shows that $V(\phi)$ is negative definite when $\phi$ approaches zero (the negativity region of $V(\phi)$ is displayed as the dotted line).}
  \label{V for lower sign}
\end{figure}

On the other hand, a non-trivial solution of Eq.  \eqref{dphi/di in terms of phi in the case Ha sinh} for the lower $``-"$ sign can be defined to be
\begin{equation} \label{phi-lower-sign}
\phi(t)=\phi_-(t)=-\sqrt{\frac{2(6+\hat\beta)}{3\hat\beta\gamma_0}}\text{arctanh} \left(M\hat\beta t\right),
\end{equation}
which implies the corresponding scale factors given by
\begin{align}\label{scale-a-lower}
a(t)&\propto \left(1 - M^2\hat\beta^2 t^2\right)^{\frac{1}{\hat\beta}},\\
\label{scale-b-lower}
b(t)&\propto \left(1 - M^2\hat\beta^2 t^2\right)^{\frac{1}{6}}.
\end{align}
It appears that $a(t)$, $b(t)$, and $\phi(t)$ are all real definite for $-1<M \hat\beta t<1$. Since $\dot{a}(t)<0$ for $0<M\hat\beta t<1$, we will only be interested in the region $-1<M\hat\beta t<0$. Similar to the above solutions, it is clear that inflation will happen if $0<\hat\beta \ll 1$. Hence, $M$ should be negative definite in this case. Interestingly, the corresponding $h(\phi)$ does not meet the strong coupling problem in this case. However, the corresponding potential $V(\phi)$ will be negative definite around $\phi=0$ provided that $0<\hat{\beta}\ll 1$ (see Fig. \ref{V for lower sign} for a demonstration). Consequently, the corresponding inflation also needs to finish before $\phi$ approaches zero.

In conclusion, we have completed the first step of the proof that is working out the corresponding forms of $V(\phi)$, $h(\phi)$, and $f(\phi)$ compatible with the constant-roll condition, $\eta_{\rm DBI}(t) =  \hat\beta$, and the associated  ones, $H_b(t)/H_a(t) = n =\hat\beta/6$ and $\gamma(t) = \gamma_0$. Up to now, we have figured out three different inflationary solutions, that are: \\
$\bullet$ solution {I}: is described by a set of Eqs. \eqref{Ha exp}, \eqref{dphi/di in terms of phi in the case Ha exp}-\eqref{scale-factor-1};\\
$\bullet$ solution {II}: is described by a set of Eqs. \eqref{Ha sinh}, \eqref{dphi/di in terms of phi in the case Ha sinh}, \eqref{explicit form of V in Ha sinh}-\eqref{explicit form of f in Ha sinh}, \eqref{phi-upper-sign}-\eqref{scale-b-upper} associated with the $\phi_+(t)$ solution;\\
$\bullet$ solution {III}:  is described  by a set of Eqs. \eqref{Ha sinh}, \eqref{dphi/di in terms of phi in the case Ha sinh}, \eqref{explicit form of V in Ha sinh}-\eqref{explicit form of f in Ha sinh}, \eqref{phi-lower-sign}-\eqref{scale-b-lower} associated with the $\phi_-(t)$ solution. 
%%%%%%%%%%%%%%%%%%%
\section{Attractor property} \label{chap4}
As mentioned above, we will prove in this section that the conditions \eqref{constant speed of sound condition}, \eqref{DBI constant-roll condition}, and \eqref{constant anisotropy condition}  are really attractive by solving numerically the field equations \eqref{DBI scalar field equation}, \eqref{1st DBI Einstein equation}, \eqref{2nd DBI Einstein equation}, and \eqref{3rd DBI Einstein equation} for three inflationary solutions {I}, {II}, and {III} with three different values $\gamma_0=1$, $1.5$, and $2$. Once numerical solutions are solved, we will plot the corresponding time evolution of $\eta_{\rm DBI}(t)$, $H_b(t)/H_a(t)$, and $\gamma(t)$ to see whether they become constant during the inflationary phase. 

It is noted that $0<\hat\beta\ll 1$ for all three inflationary solutions.  Additionally, it is also noted that $M >0$ for both solutions {I} and {II}, while $M<0$ for the solution {III}. Therefore, we will choose $\hat\beta=0.1$, $|M|=10^{-4}$, along with initial conditions for scalar field as $\phi(0)=15$ and $\dot\phi(0)=0$,
 for all three solutions {I}, {II}, and {III}. By choosing that, it is easy to compare these solutions to each other. It is noted that these values are similar to that used in the canonical case \cite{Ito:2017bnn} since we would like to see the effect of $\gamma_0$. 
 
\begin{figure}[hbtp]
\includegraphics[scale=0.35]{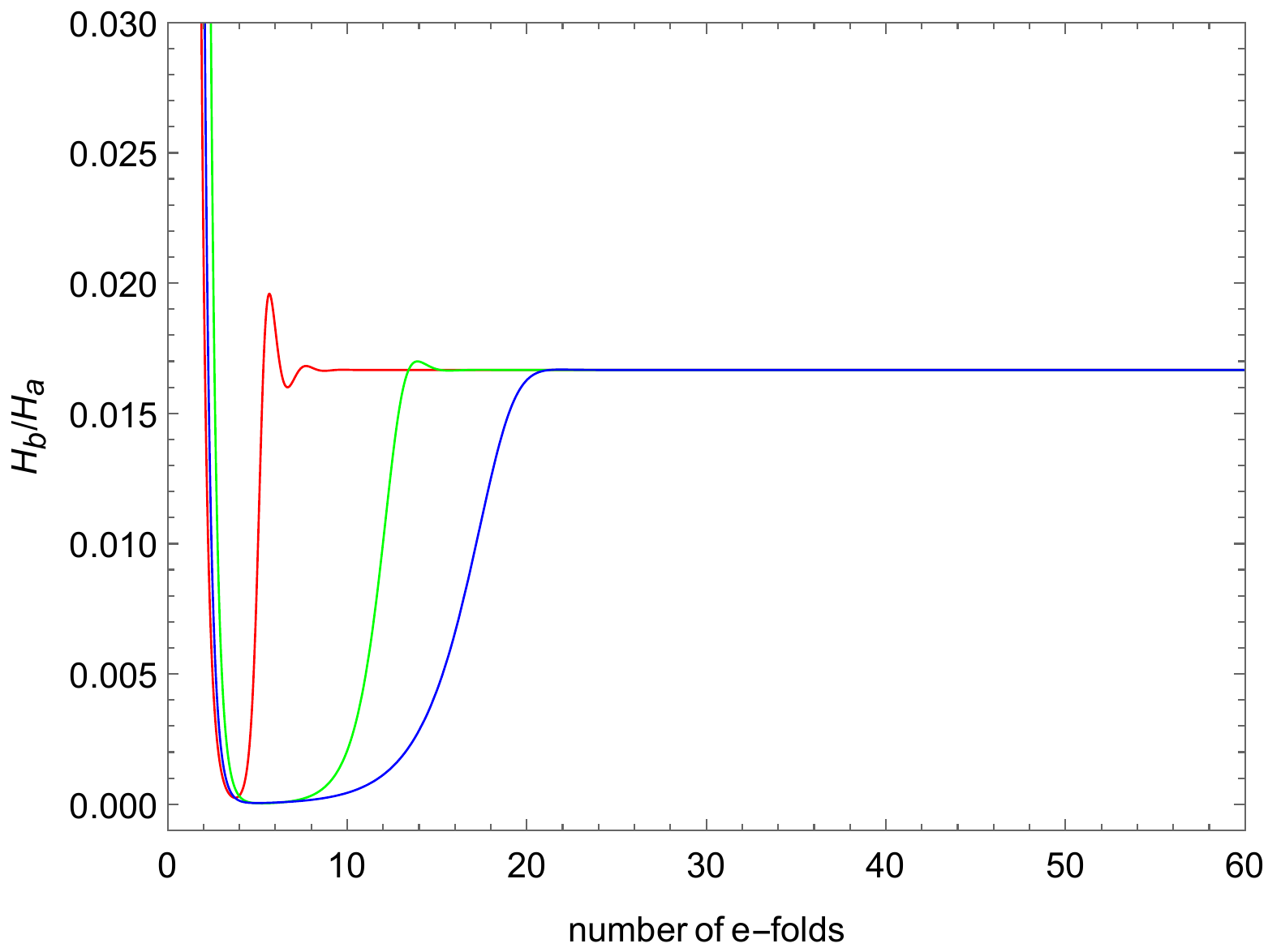}\quad
\includegraphics[scale=0.35]{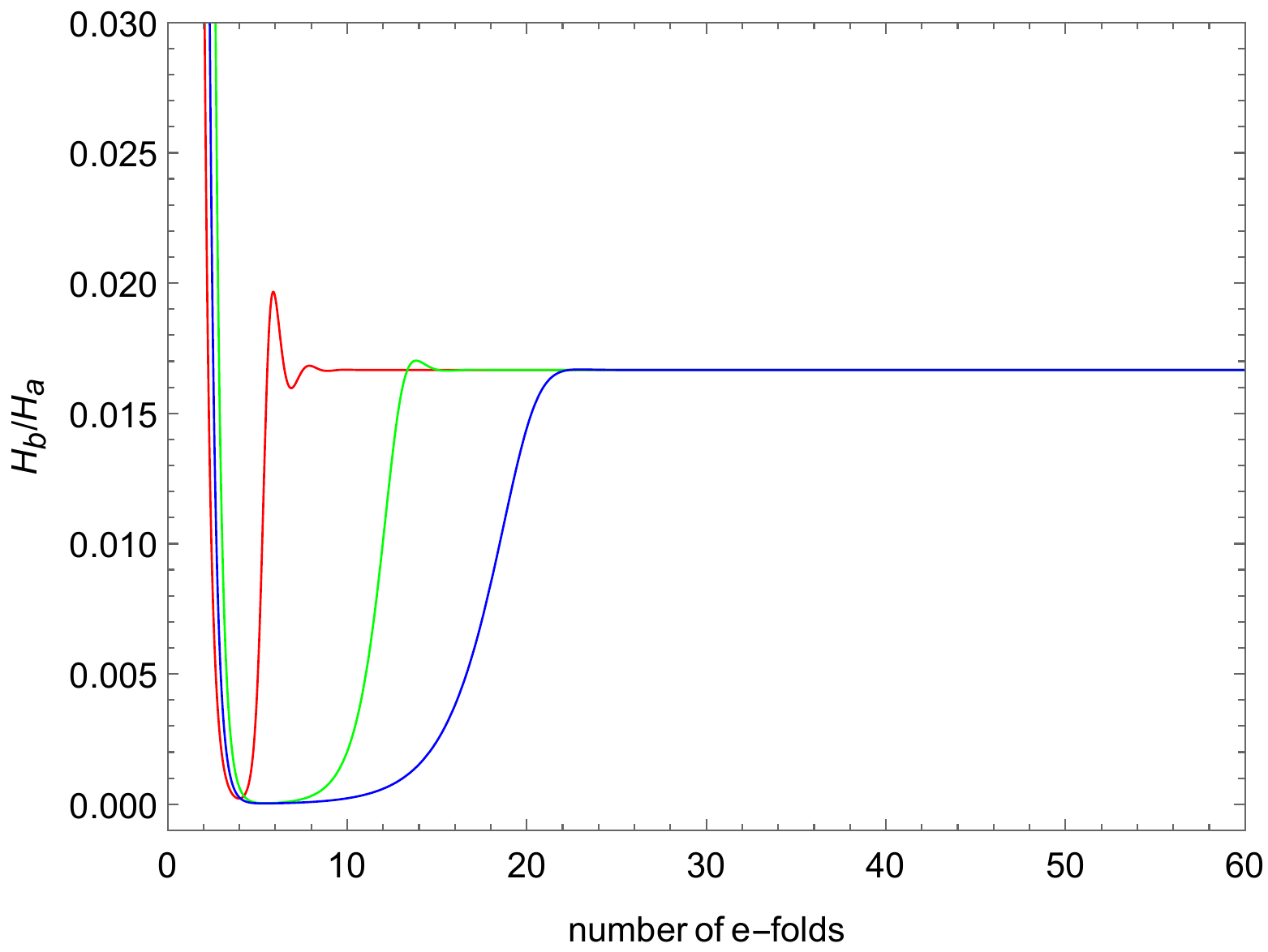}\quad
\includegraphics[scale=0.35]{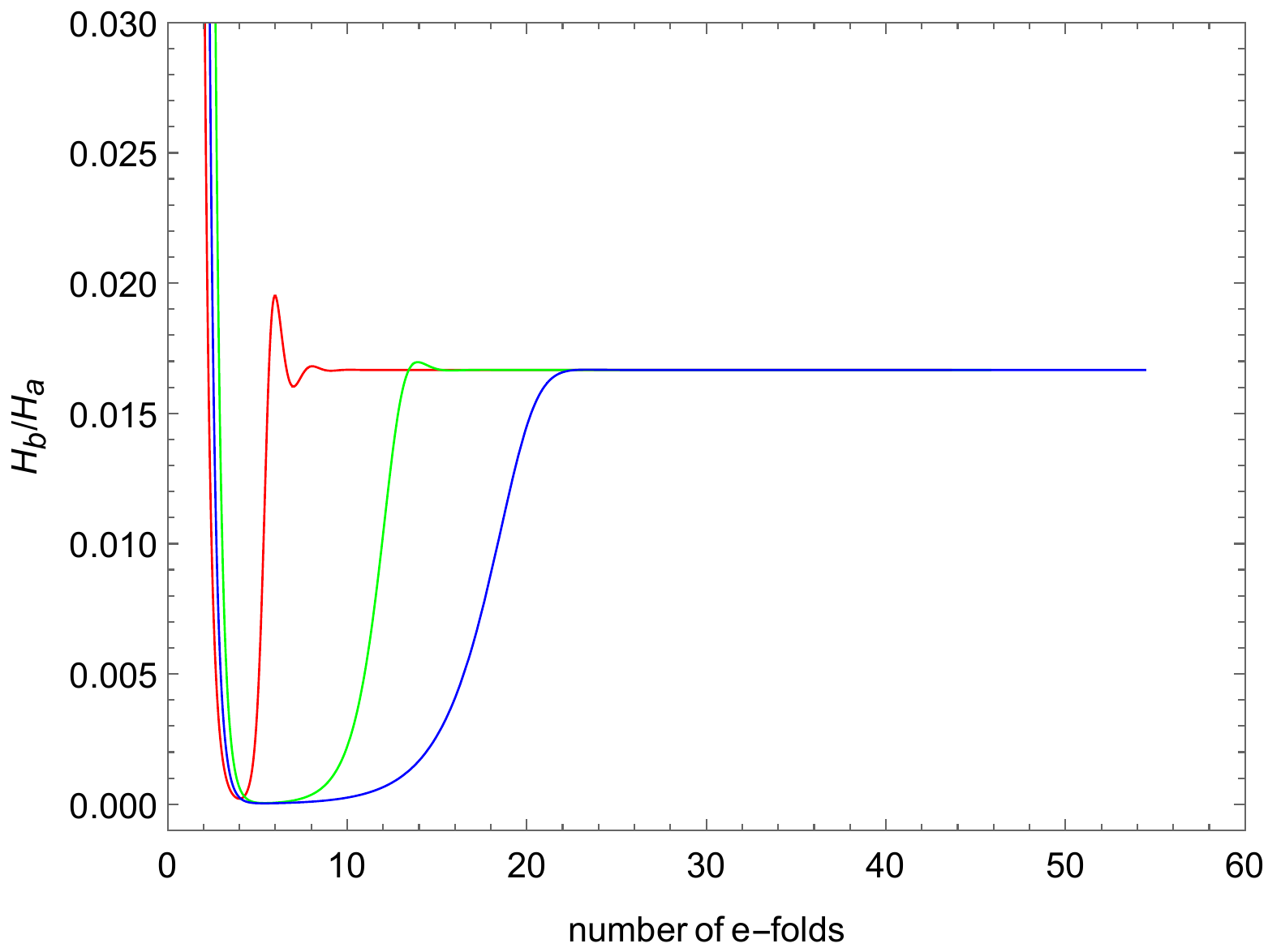}\\
\centering
\caption{Time evolution of the ratio $H_b(t)/H_a(t)$ for three different values of $\gamma_0$. These plots clearly show that all trajectories converge to the same constant, i.e., $H_b(t)/H_a(t) \to n=\hat\beta/6 \simeq 0.0167$, confirming the attractive property of the solution $n=\hat\beta/6$. The left, middle, and right plots correspond to the solutions {I}, {II}, and {III}, respectively. The red, green, and blue curves correspond to $\gamma_0=1$, $1.5$, and $2$, respectively.}
\label{evolution of n}
\end{figure}

The time evolution of $H_b(t)/H_a(t)$ is depicted in Fig. \ref{evolution of n}. It is clear that this ratio converges to the assumed constant, $n=\hat\beta/6 \simeq 0.0167$, after $10-20$ e-folds. On the other hand, the value of $\gamma_0$ does affect on the evolution of $H_b(t)/H_a(t)$. In particular, the increase of the value of $\gamma_0$ tends to delay the convergence of $H_b(t)/H_a(t)$.

\begin{figure}[hbtp]
\includegraphics[scale=0.35]{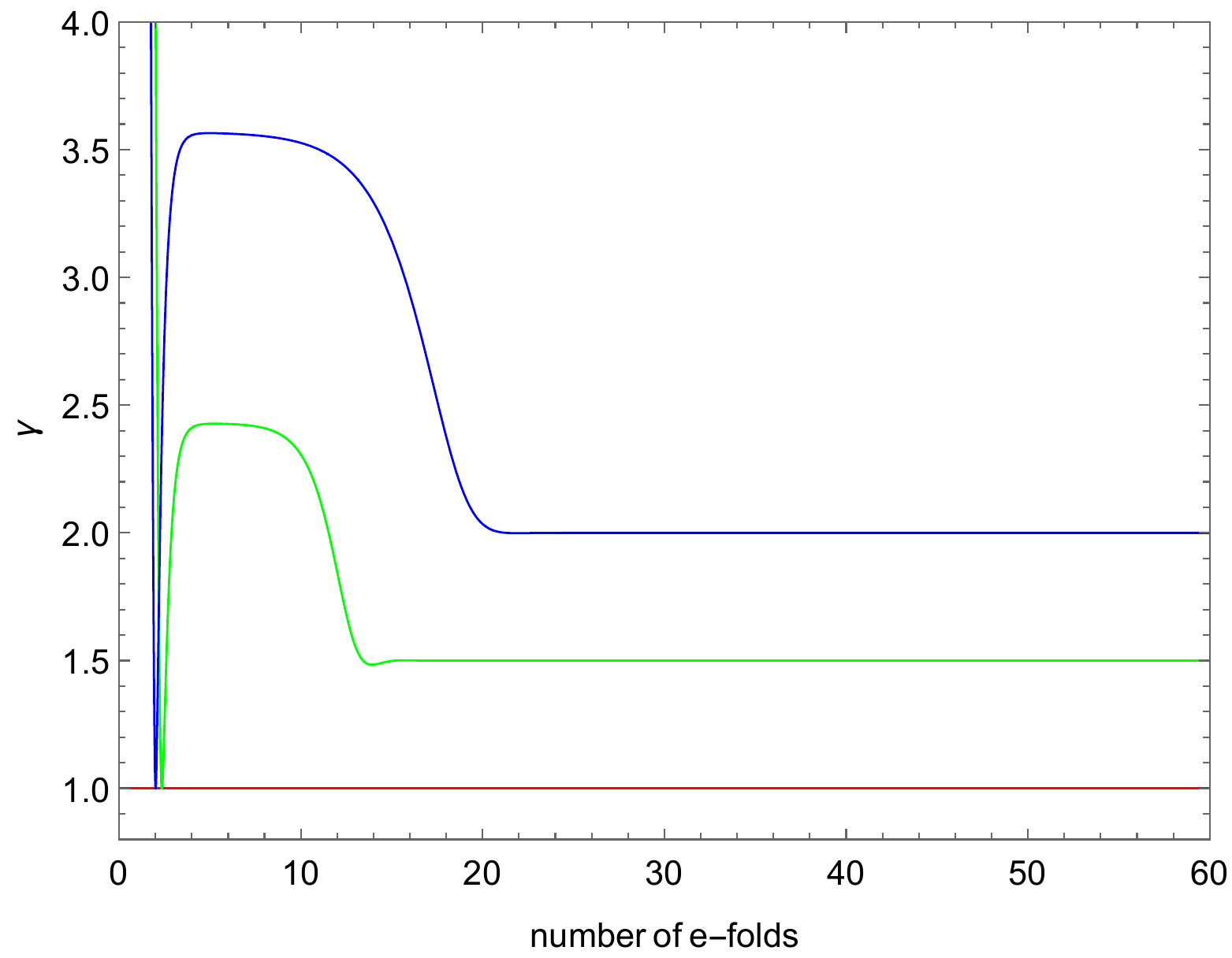}\quad
\includegraphics[scale=0.35]{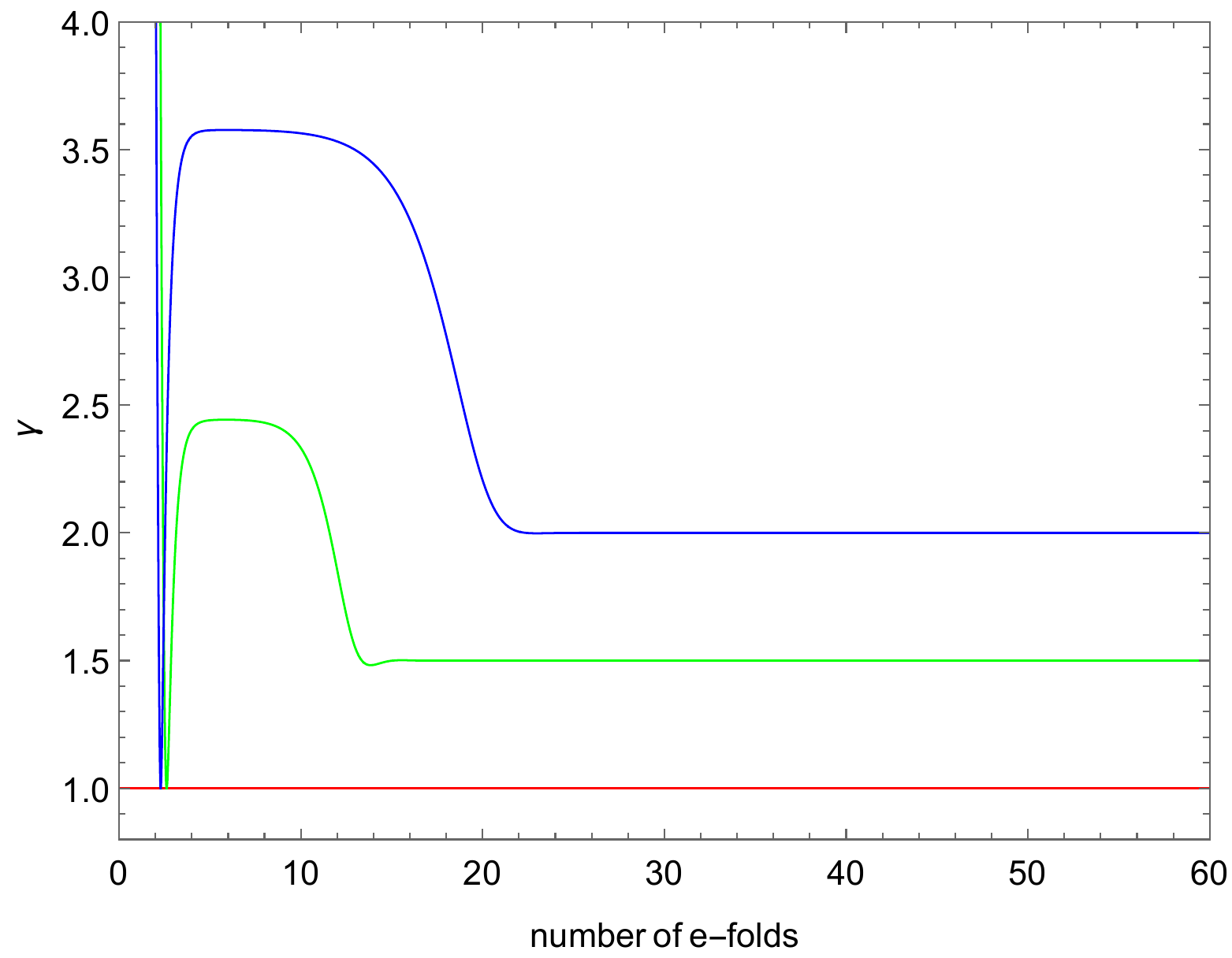}\quad
\includegraphics[scale=0.35]{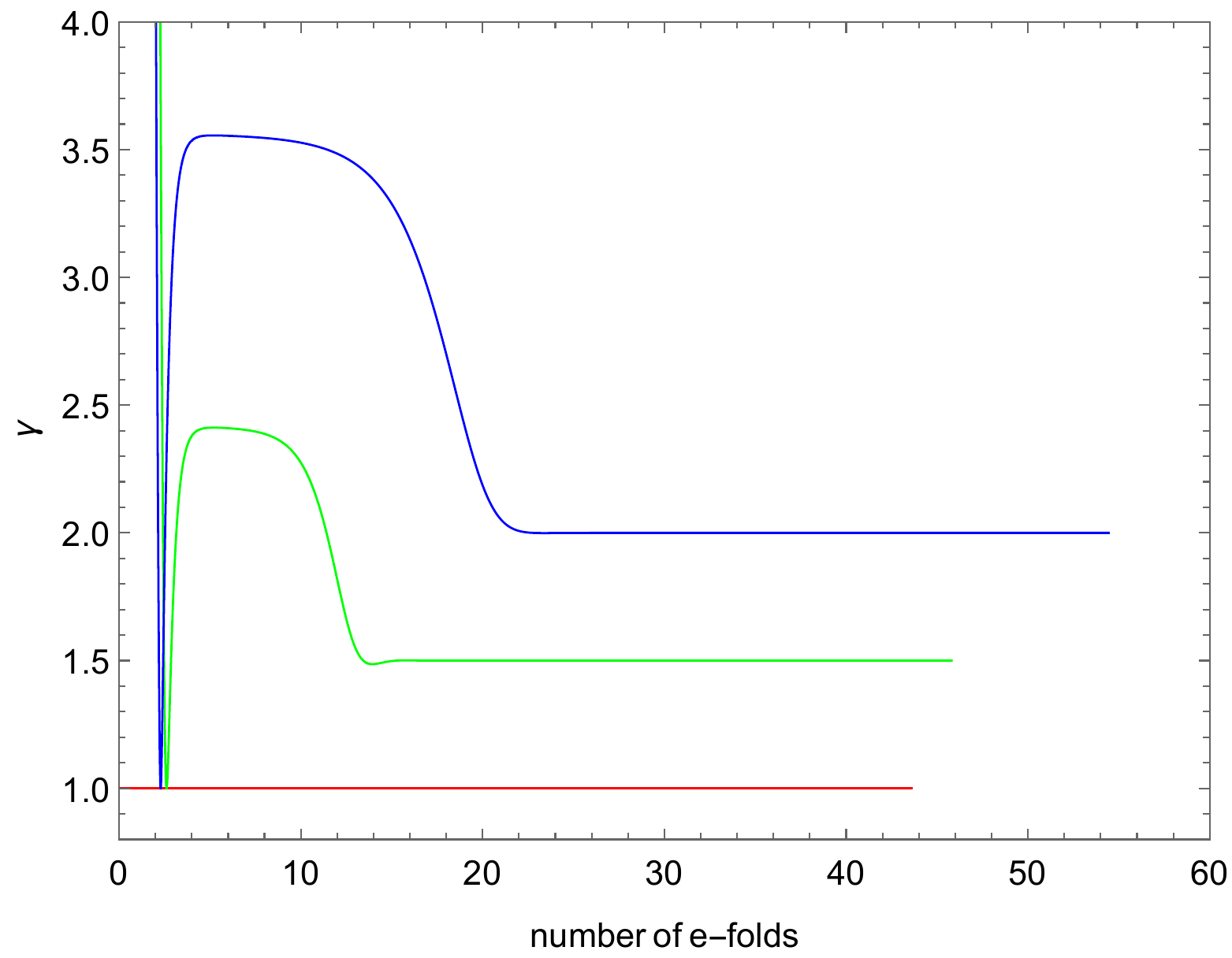}\\
\centering
\caption{ Time evolution of $\gamma(t)$ for three different values of $\gamma_0$. These plots clearly show that all $\gamma(t)$ converge exactly to their corresponding $\gamma_0$, confirming the attractive property of the solution $\gamma(t) = \gamma_0$. The left, middle, and right plots correspond to the solutions {I}, {II}, and {III}, respectively. The red, green, and blue curves correspond to $\gamma_0=1$, $1.5$, and $2$, respectively.}
\label{evolution of gamma}
\end{figure}

The time evolution of $\gamma(t) = 1/\sqrt{1-f(\phi)\dot{\phi}^2}$ is displayed in Fig. \ref{evolution of gamma}. It is noted that $\gamma(t)$ should always be larger than or equal to one. For $\gamma_0=1$, $\gamma(t)$ should remain as one constantly as time evolves. As a result, these properties can be easily confirmed by the plots in Fig. \ref{evolution of gamma}. Additionally, these plots also clearly indicate that $\gamma(t) \to \gamma_0 $ after $10-20$ e-folds. More interestingly, it is shown that the value of $\gamma_0$ also affects on the evolution of $\gamma (t)$ in the same way it does for $H_b(t)/H_a(t)$. These results confirm that the condition $\gamma(t)=\gamma_0$ as chosen in Eq. \eqref{constant speed of sound condition} is really attractive. 

\begin{figure}[hbtp]
 {\includegraphics[scale=0.35]{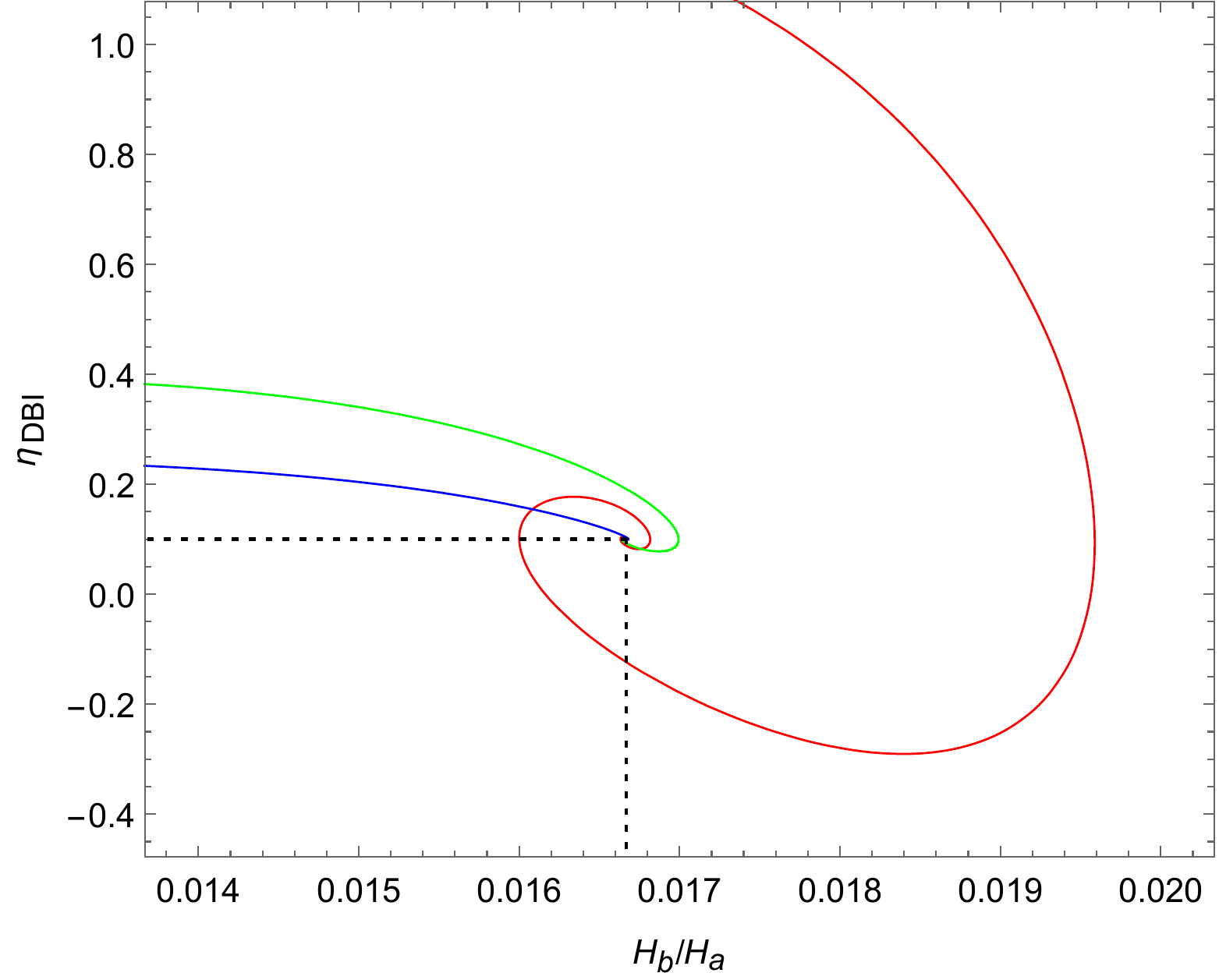}}\quad
 {\includegraphics[scale=0.35]{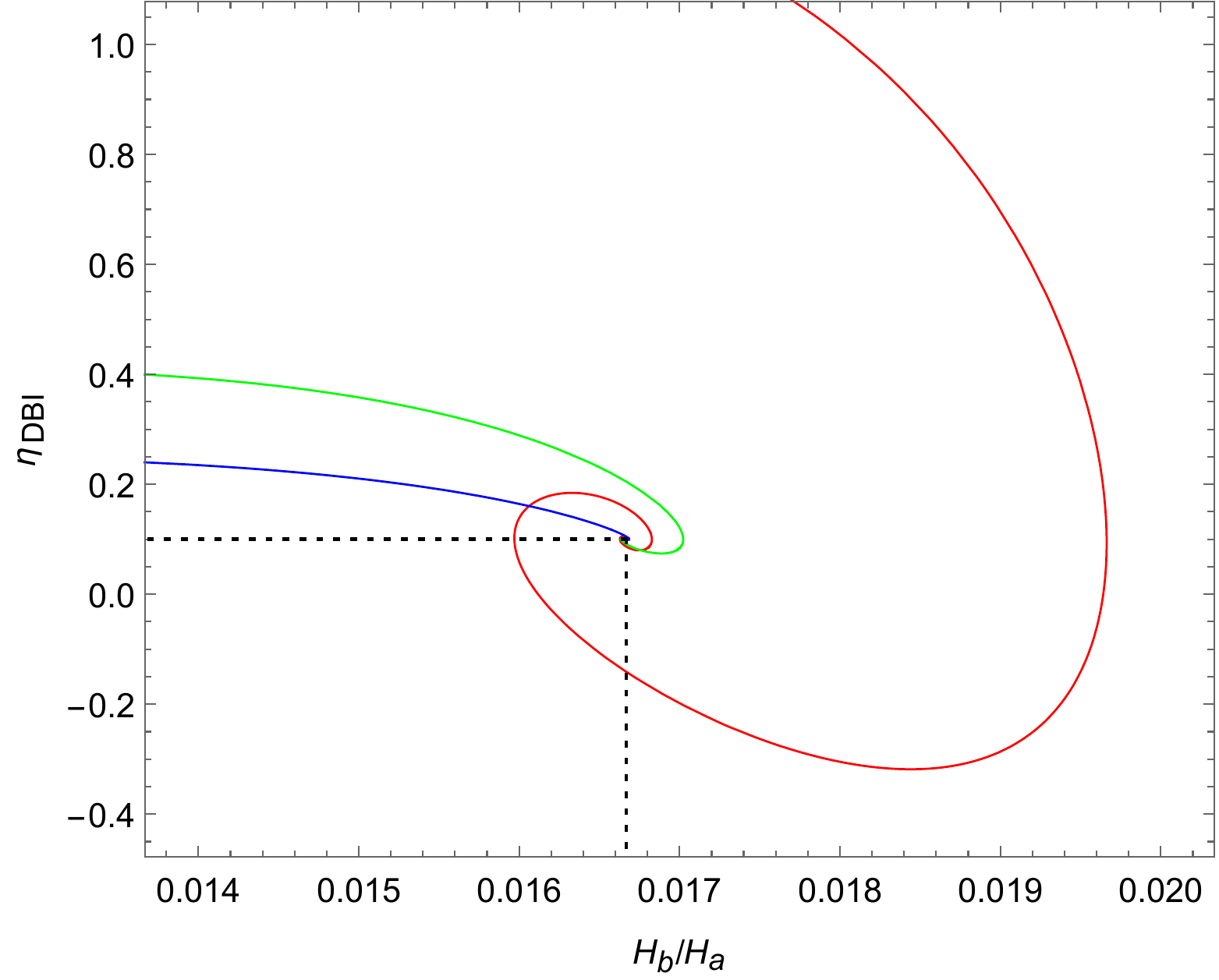}}\quad
  {\includegraphics[scale=0.35]{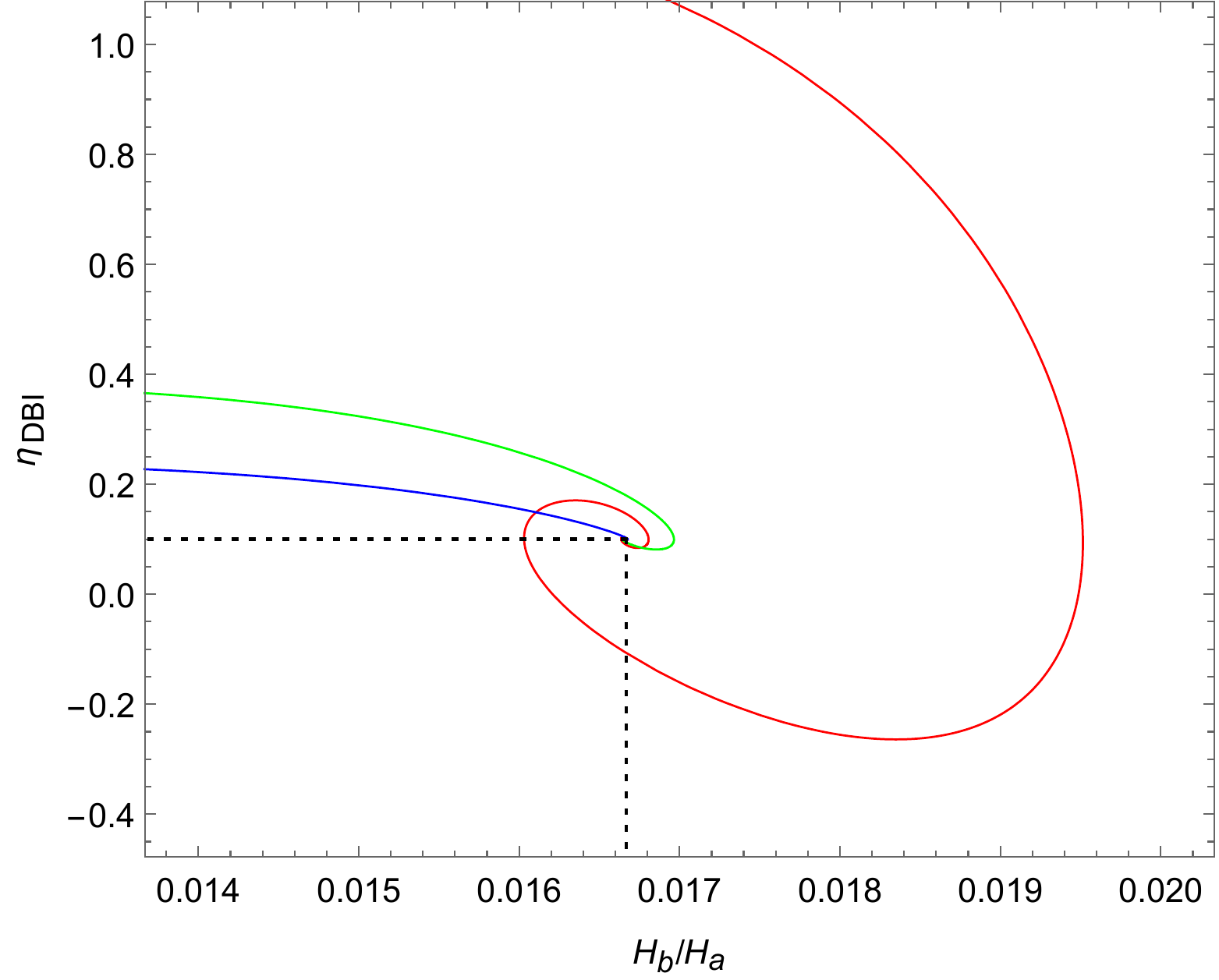}}\\
   \centering
 \caption{Phase space of $H_b(t)/H_a(t)$ and $\eta_{\rm DBI}(t)$ for three different values of $\gamma_0$. The left, middle, and right plots correspond to the solutions {I}, {II}, and {III}, respectively. The red, green, and blue curves correspond to $\gamma_0=1$, $1.5$, and $2$, respectively.}
  \label{phase space of n and beta}
\end{figure}

Next, we plot in Fig. \ref{phase space of n and beta} the phase space of $H_b(t)/H_a(t)$ and $\eta_{\rm DBI}(t)$. We see that for three different values of $\gamma_0$, the trajectories all converge to the same fixed point $\left(\hat\beta,n \right) \simeq \left(0.1,0.0167 \right)$.

\begin{figure}[hbtp]
 {\includegraphics[scale=0.45]{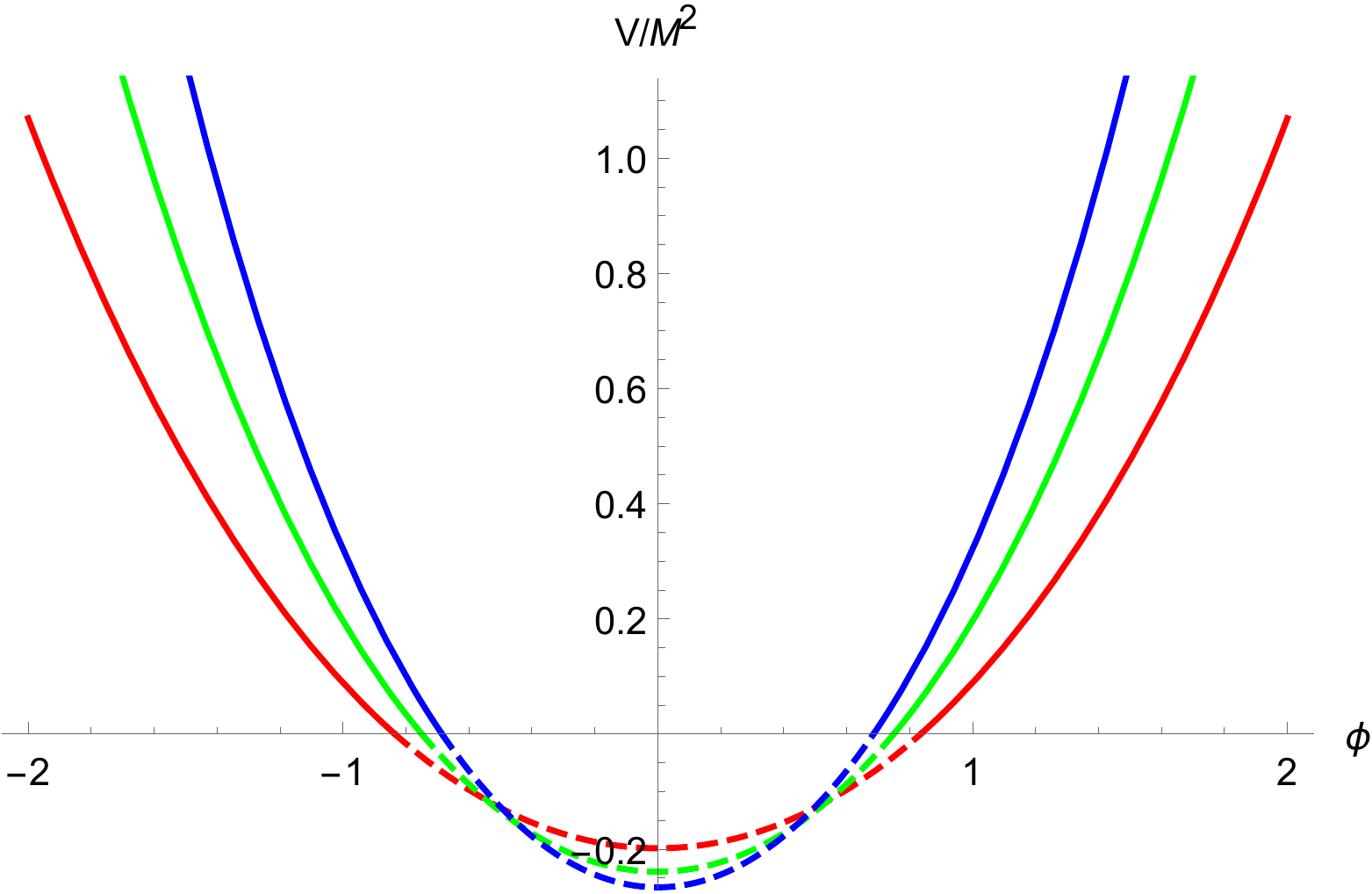}}\\
   \centering
 \caption{Behavior of potential $V(\phi)$ associated with the solution {III} for different values of $\gamma_0$. The red, green, and blue curves correspond to $\gamma_0=1$, $1.5$, and $2$, respectively.}
  \label{V_all}
\end{figure}

Finally, we would like to discuss the difference between three solutions {I}, {II}, and {III}. According to the plots, it turns out that the behavior of these solutions are not significantly different from each other. The only gap between the first two solutions {I} and {II} and the last solution {III} is the e-fold number as displayed in Fig. \ref{evolution of gamma}. In particular, the inflationary phases corresponding to the first two solutions {I} and {II} still survive when the e-fold number reaches to 60.  However, this is not the case for the last solution {III}. Indeed, the corresponding inflationary phase of this solution stops before the e-fold number reaches to 60 due to the negativity of the corresponding potential $V(\phi)$ happening when $\phi$ approaches zero as mentioned above. One additional interesting point should be noted that the value of $\gamma_0$ does affect on the e-fold number of the inflationary solution {III}. As a result, it turns out that the larger $\gamma_0$ is, the larger e-fold number the inflationary solution {III} has. This result can be explained according to the Fig. \ref{V_all}. In particular, this figure clearly displays that the potential $V(\phi)$ tends to become negative sooner when $\gamma_0$ decreases to one. 

%%%%%%%%%%%%%%%
\section{Conclusions} \label{final}
In this paper, we have extended our DBI model proposed in Ref. \cite{DBI}, which is a non-canonical extension of the KSW model \cite{MW0}, to a novel type of inflation called the constant-roll inflation \cite{Motohashi:2014ppa}. Basically, our paper is just a non-trivial extension of the paper \cite{Ito:2017bnn}, in which anisotropic constant-roll inflationary solutions have been found to the KSW model. As a result, we have figured out three exact anisotropic constant-roll inflationary solutions named as  I, {II}, and {III}, to the DBI model. As a result, these solutions have been achieved under the constant-roll condition \cite{Motohashi:2014ppa,Mohammadi:2018zkf},  $\eta_{\rm DBI} (t) =  \hat\beta$, along with the compatible ones, $H_b(t)/H_a(t) = n=\hat\beta/6$ for the Bianchi type I spacetime \cite{Ito:2017bnn} and $\gamma(t) = \gamma_0$ for the DBI field. While the solution {I} is nothing but the well-known power-law type \cite{DBI}, the other ones {II} and {III} turn out to be novel ones. Numerical calculations have been performed to confirm that these conditions are really attractive. These results together with that investigated in Ref. \cite{Ito:2017bnn}  indicate that the cosmic no-hair conjecture is extensively violated in the KSW model as well as in its non-canonical extension, the DBI model, even when the constant-roll condition is applied. It would be interesting if ones could find anisotropic constant-roll inflationary solutions to other non-canonical models such as the Galileon inflation and the $k$-inflation \cite{SDBI,ghost-condensed,Odintsov:2019ahz} as well as to multi-field models \cite{WFK,Guerrero:2020lng}. Additionally, it is important to investigate the CMB imprints of the anisotropic constant-roll inflation. For example, one might ask whether this model satisfies the observational constraints on the spectral index and tensor-to-scalar ratio. However, it turns out that these observational issues are not straightforward to answer within not only the framework of constant-roll inflation \cite{Motohashi:2017aob,GalvezGhersi:2018haa,Mohammadi:2018zkf} but also the framework of anisotropic inflation \cite{Imprint2,Imprint3,Imprint4}.  We will therefore leave them to our further studies, whose results will be presented in a sequel to this paper. We hope that our present paper would be useful to studies of anisotropic inflation as well as constant-roll inflation. 
%%%%%%%%%%%%%%%
\begin{acknowledgments}
We would like to thank the referee very much for useful comments and suggestions. D.H.N. would like to thank Dr. Asuka Ito very much for his useful discussions. T.Q.D. would like to thank Prof. W. F. Kao very much for his fruitful collaborations on the previous works of anisotropic inflation. D.H.N. was funded by Vingroup Joint Stock Company and supported by the Domestic Master/ PhD Scholarship Programme of Vingroup Innovation Foundation (VINIF), Vingroup Big Data Institute (VINBIGDATA), code VINIF.2020.ThS.65. This study is supported by the Vietnam National Foundation for Science and Technology Development (NAFOSTED) under Grant number 103.01-2020.15.  
\end{acknowledgments}
%%%%%%%%%%%%%%%%%

\end{document}